# Pericentric passage-driven star formation in satellite galaxies and their hosts: CLUES from Local Group simulations

Arianna Di Cintio[1,2⋆], Robert Mostoghiu [3], Alexander Knebe[3,4,5] & Julio F. Navarro[6]

[1] *Universidad de La Laguna. Avda. Astrofísico Fco. Sánchez, La Laguna, Tenerife, Spain*
[2] *Instituto de Astrofísica de Canarias, Calle Via Láctea s/n, E-38206 La Laguna, Tenerife, Spain*
[3] *Departamento de Física Teórica, Módulo 15, Facultad de Ciencias, Universidad Autónoma de Madrid, 28049 Madrid, Spain*
[4] *Centro de Investigación Avanzada en Física Fundamental (CIAFF), Facultad de Ciencias, Universidad Autónoma de Madrid, 28049 Madrid, Spain*
[5] *International Centre for Radio Astronomy Research, University of Western Australia, 35 Stirling Highway, Crawley, Western Australia 6009, Australia*
[6] *Department of Physics and Astronomy, University of Victoria, Victoria, BC V8P 5C2, Canada*

## ABSTRACT

Local Group satellite galaxies show a wide diversity of star formation histories (SFHs) whose origin is yet to be fully understood. Using hydrodynamical simulations from the Constrained Local UniversE project, we study the SFHs of satellites of Milky Way-like galaxies in a cosmological context: while in the majority of the cases the accretion onto their host galaxy causes the satellites to lose their gas, with a subsequent suppression in SF, in about 25% of our sample we observe a clear enhancement of SF after infall. Peaks in SF clearly correlate with the satellite pericentric passage around its host and, in one case, with a satellite-satellite interaction. We identify two key ingredients that result in enhanced SF after infall: galaxies must enter the host's virial radius with a reservoir of cold gas $M_{\rm gas,inf}/M_{\rm vir,inf} \gtrsim 10^{-2}$ and with a minimum pericentric distance $\gtrsim$10 kpc (mean distance ∼50 kpc for the full sample), in order to form new stars due to compression of cold gas at pericentric passage. On the other hand, satellites that infall with little gas or whose pericentric distance is too small, have their gas ram-pressure stripped and subsequent SF quenched. The pericentric passage of satellites likewise correlates with SF peaks in their hosts, suggesting that this mechanism induces bursts of SF in satellites and central galaxies alike, in agreement with recent studies of our Galaxy's SFH. Our findings can explain the recently reported multiple stellar populations observed in dwarf galaxies such as Carina and Fornax, and should be taken into account in semi-analytic models of galaxy formation and satellite quenching.

## 1 INTRODUCTION

Studying dwarf galaxies and in particular understanding their star formation histories (SFHs) is crucial for our comprehension of their formation and evolution within a cosmological context. The Local Group (LG) of galaxies is the ideal area in the Universe to perform such study, given the vicinity of its dwarf galaxy population to our own Milky Way (e.g. Mateo 1998, and references therein). Detailed SFHs can be obtained by means of accurately modelling colour-magnitude diagrams in combination with the ability of resolving individual stars, using the Hubble Space Telescope (HST) (e.g. Weisz et al. 2011, 2014; Skillman et al. 2017) and more recently thanks to the newly delivered Gaia DR2 data (e.g. Gallart et al. 2019; Fritz et al. 2020; McConnachie & Venn 2020). A large body of observational work suggests that LG dwarfs have notably diverse SFHs: understanding such variety and their origin is at the very center of current observational and theoretical efforts (Dolphin et al. 2005; Benítez-Llambay et al. 2015; Gallart et al. 2015 and references therein).

Several mechanisms can affect the star formation of dwarf galaxies across cosmic times, amongst which we note the prominent role of reionization, which is expected to evaporate gas from the haloes of small galaxies (Shapiro et al. 1994), the interactions with large-scale filaments of gas, the so-called 'cosmic web strip-ping' (Benítez-Llambay et al. 2013), that is able to remove gas from dwarfs, thus preventing further star formation, and the role of internal stellar feedback, that expels the gas via energetic outflow episodes (Governato et al. 2010).

⋆ Junior Leader Fellow 'La Caixa' Foundation, adicintio@iac.es

In Benítez-Llambay et al. (2015) the authors show that the main quantities that determine the SFH of isolated dwarf galaxies are the halo mass of each galaxy at reionization time and their subsequent mass accretion history. The combination of these factors leads to different star formation times that can be summarized as continuous, old, and old plus young SFHs. Galaxies that are already massive at redshift z=6 are sufficiently large to retain their gas during reionization and continue to form stars, showing a declining SFH with time as more gas is consumed. On the contrary, galaxies with a total mass, including dark matter, of less than $10^{8.5}\mathrm{M}_\odot$ at reionization, have all their gas photo-evaporated and can only form stars before z=6, ending up with a single, old stellar populations.

In addition to these mechanisms, we should consider those processes that are specific of dwarf satellite galaxies, such as ram pressure stripping, as the satellites pass through the hot cloud of gas around their host, leading to a truncation or reduction of their star formation, and tidal stripping and 'stirring', as an alternative way of removing gas from dwarf galaxies and that leads to fundamental changes in their configuration (Mayer et al. 2006). All of these physical processes can be at work simultaneously in satellite galaxies of the LG, to give rise to the current variety of SFHs that we observe in dwarf spheroidals (dSphs) around our Galaxy, as indicated by deep photometric observations of Galactic satellites with HST (Weisz et al. 2014).

While all dSphs contain an ancient stellar population, which in most cases is the dominant component (e.g., Sculptor, Draco, Ursa Minor), in other cases an intermediate age stellar component is also present (e.g., Carina, Fornax, Leo I). The origin of such multiple populations in dwarf galaxies such as Carina, Leo I or Fornax remains an open question (Hurley-Keller et al. 1998; Gallart et al. 1999; Monelli et al. 2003; Bono et al. 2010; Monelli et al. 2014; del Pino et al. 2013). In particular, the different age-metallicity relations derived for the two main SF peaks of the Carina dSphs, occurred at old and intermediate ages, clearly indicate that such stars formed from gas with different abundance patterns, which is inconsistent with a simple evolution in isolation of this dwarf (de Boer et al. 2014; Hayashi et al. 2018).

Previous observational work has suggested that the close passage of a satellite galaxy near its host could be connected with a peak in its SFH (e.g. Sohn et al. 2007; Pasetto et al. 2011; Rocha et al. 2012): such early findings have been recently supported by the work of Rusakov et al. (2021), who find a correspondence between the main intermediate-age and young SFH events of Fornax dSphs and its pericentric passages around the Milky Way. Moreover, it has been shown in novel work that the SFH of the host galaxy itself could be impacted by the pericentric passage of its satellites: three narrow episodes of enhanced star formation have been derived for the MW, whose timing coincide very well with the proposed Sagittarius dwarf galaxy pericentric passage (Ruiz-Lara et al. 2020).

Some early simulations hinted to a link between the surface density profile of gas bound to the dwarfs, which steepens remarkably at each pericenter passage because of tidal compression and torques (Mayer et al. 2001). Other literature based on hydrodynamical cosmological simulations suggests that, for isolated galaxies, star formation can be reignited due to interactions with streams of gas in the intergalactic medium (Wright et al. 2019) or by further accretion of gas that may rekindle star formation even in galaxies quenched by cosmic reionization (Rey et al. 2020), while dwarf-dwarf mergers can offer a viable scenario for the formation of galaxies with multiple distinct populations (Benítez-Llambay et al. 2016).

More recently, Miyoshi & Chiba (2020) considered a time-varying gravitational potential for the MW, to calculate the orbits of Galactic dSphs, guided with Gaia DR2 proper motions, and found that the infall time of a satellite coincides well with the time when the star formation rate (SFR) peaked for classical dSphs. Fillingham et al. (2019) used DM only simulations to infer quenching timescales for low-mass satellites around the MW, showing a rapid cessation of star formation following infall and quenching timescales that are shorter for those dSphs having high orbital eccentricities; they further note that Carina and Fornax are on orbits with relatively large pericenters of 60 and 58 kpc, respectively. Simpson et al. (2018) used Auriga simulations to show that ram pressure stripping appears to be the dominant quenching mechanism for satellites, with 50% of quenched systems ceasing star formation within 1 Gyr of first infall; furthermore, they show a compression of gas within the satellite at pericentric passage, resulting in a small enhancement in star formation. Hausammann et al. (2019), additionally, showed that while ram pressure is very efficient at stripping the hot and diffuse gas around dwarfs, the cold gas is more resilient to stripping and, under the right conditions, may be able to prolong star formation in satellites.

Genina et al. (2019) investigated the mechanisms that lead to the formation of two spatially segregated metallicity populations in Local Group dwarf galaxies: they found that if a satellite galaxy is able to retain some of its gas at infall, it is then the compression of such gas, by ram pressure near pericentre, that triggers the formation of new, metal-rich stars. Furthermore, the role of a satellite's orbit in explaining the kinematical separation between metal poor and metal rich stars was explored in Sculptor and Carina dSphs in Sales et al. (2010). Finally, Applebaum et al. (2021) recently ran ultra high-resolution simulations of ultra-faint dwarfs to show that most of them are quenched prior to interactions with the Milky Way, though some can retain their gas until infall.

Aside from simulations, semi-analytic models have historically implemented schemes in which once a galaxy becomes a satellite it stops accreting gas, leading to its subsequent quenching (e.g. Starkenburg et al. 2013, and references therein). On top of that, tidal and ram-pressure stripping act on the already existing gas, with only a few semi-analytic models accounting simultaneously for ram-pressure stripping of cold gas as well as a gradual stripping of hot gas (Stevens & Brown 2017; Cora et al. 2018; Xie et al. 2020). Regardless of the specifics of each semi-analytic model, the so-far implemented processes tend to remove the fuel for star formation in satellites, such that the modelled satellites are often systematically gas-poor (Stevens & Brown 2017).

As such, there is mounting evidence that the orbital configuration as well as the gas availability at the moment of infall play a fundamental role in shaping the evolution and star formation history of satellite galaxies, such as the Local Group ones.

Motivated by previous observational and theoretical work, in particular by the one of Miyoshi & Chiba (2020) and Rusakov et al. (2021), we investigated the origin of the diverse SFHs of satellite galaxies within hydrodynamical cosmological simulations of the Local Group from the CLUES project (Gottlöber et al. 2010; Carlesi et al. 2016). The advantage of these constrained simulations is that they offer a unique opportunity to perform studies of LG satellites, since they reproduce faithfully the LG environment.

The aim of our work is to understand the role of the environment in shaping the SFHs of satellite galaxies, and in particular to explore under what conditions a satellite's infall into the main host leads to either quenching or enhancing of its star formation.

This paper is organized as follows: in Section 2 we describe the hydrodynamical cosmological simulation of the Local Group used for this study; in Section 3.1 we present our sample of satellite galaxies, compare them with a control sample of simulated isolated galaxies in Section 3.2 and quantify the role of infall in Section 3.3. We then proceed to discuss the main results of this work: we study in detail the differences between quenched and enhanced SFHs arising after infall in Section 4.1, in Section 4.2 we demonstrate that the necessary conditions to develop star formation in a satellite after infall depend on gas availability and pericentric distances, and we show that this can influence the SFH of the host galaxy itself in Section 4.3. Finally, we compare our simulations with recent literature data from the Gaia satellite in Section 4.4, before concluding in Section 5.

## 2 SIMULATIONS

In order to explore differences in star formation histories between satellite and isolated galaxies, we use one of the Constrained Local Group Simulations from the CLUES project (Gottlöber et al. 2010).[1] The particular simulation used here is called `Clues3_LGGas` and has already been extensively analyzed elsewhere (e.g. Libeskind et al. 2010, 2011, 2013; Knebe et al. 2010, 2011; Di Cintio et al. 2011, 2013); the halo catalogues and merger trees are, further, publicly available.[2] While all the physical and technical information can be found in the aforementioned papers and the database website, respectively, we include a brief summary here for completeness.

The CLUES simulations are performed with the treePM N-body + SPH code GADGET2 [3] (Springel 2005), and their initial conditions are constrained such that the observed Local Volume, on scales smaller than $\approx 5\,h^{-1}$ Mpc, is reproduced. A zoom-in, higher resolution simulation is performed in a region of $2h^{-1}$ Mpc centered on the Local Group (LG) with $4096^3$ effective particles within it. Within such region, it is possible to identify the three main galaxies formally corresponding to the Milky Way (MW), Andromeda (M31) and Triangulum (M33) galaxies. The simulations assume a WMAP3 cosmology (Spergel et al. 2007), i.e. $\Omega_m = 0.24$, $\Omega_b = 0.042$, $\Omega_\Lambda = 0.76$ and $h = 0.73$, a normalization of $\sigma_8 = 0.75$ and a slope of the power spectrum of $n = 0.95$. The particle mass resolution is $m_{\rm DM} = 2.1 \times 10^5 h^{-1}{\rm M}_\odot$ for the dark matter particles, $m_{\rm gas} = 4.4 \times 10^4 h^{-1}{\rm M}_\odot$ for the gas particles and $m_{\rm star} = 2.2 \times 10^4 h^{-1}{\rm M}_\odot$ for the star particles. The gravitational softening length is $\epsilon = 150 h^{-1}$pc.

The Amiga-Halo-Finder AHF[4] (Knollmann & Knebe 2009; Gill et al. 2004) has been used in combination with its bundled MERGERTREE code to identify and track in time all haloes and subhaloes in the simulation. The (virial) mass of each halo is defined as the mass of a sphere containing $\Delta = 94$ times the critical matter density of the Universe $\rho_{\rm crit} = 3H^2/8\pi G$ at z=0, unless the density profiles rises before, which might be the case for subhaloes (for which the radius is then the upturn point). In any case we will refer to the (sub-)halo mass as $M_{\rm vir}$ and its radius as $R_{\rm vir}$.

The feedback and star formation recipes are described in Springel & Hernquist (2003). Essentially, the interstellar medium (ISM) is modelled as a two-phase medium composed of hot ambient gas and cold gas clouds in pressure equilibrium. The thermodynamic properties of the gas are computed in the presence of a uniform but evolving ultra-violet cosmic background switched on at z=6 (Haardt & Madau 1996). Cooling rates are calculated from a mixture of a primordial plasma composition and no metal dependent cooling is assumed. Star formation is treated stochastically, in order to reproduce the Kennicutt-Schmidt law for spiral galaxies (Schmidt 1959; Kennicutt 1998). The first time a gas particle undergoes star formation, a star particle of half the gas particle's mass is created, reducing the gas particle mass appropriately. The second episode of star formation results in the gas particle converting all its remaining mass into a star particle. Thus all star particles have the same mass of $m_{\rm STAR} = 2.21 \times 10^4 h^{-1}{\rm M}_\odot$, while gas particles have either their full original mass, or half their original mass.

## 3 ANALYSIS

### 3.1 Satellite Galaxy Sample

We select satellites belonging to one of the three main galaxies in the simulated LG, namely the Milky Way, Andromeda or the Triangulum galaxy. All satellites used here are chosen to be more massive than $M_{\rm vir} = 10^9{\rm M}_\odot$ at infall time, and to contain at least 100 star particles at redshift z=0: this selection criteria provides a sample of 23 objects in total. The star formation histories are then inferred from the age histogram of the bound stars found within each satellite galaxy at z=0.

In Fig. 1 the star formation histories of the 23 satellites in our sample are shown, together with their stellar, gas and halo mass at z=0 and satellite's ID (unique identifier from the halo finder catalogue[5]). SFHs are normalized to their peak value, and computed in 50 linearly-spaced bins between 0 and 13.73 Gyrs, so that the bin-width is about 0.28 Gyrs. The first 12 objects are satellites of M31, the following seven are satellites of the MW and the last four are satellites of M33. The vertical dashed red line in each panel indicates the time of first infall of each of these objects into their respective host, defined as the time at which the satellite first crosses its host's virial radius. Some of the satellites may have "backsplashed" (Balogh et al. 2000; Gill et al. 2005; Sales et al. 2007), and in that case the first infall is considered as the relevant one. Fig. 1 shows the large variety of SFHs of our simulated satellites, not unlike what is observed in the Local Group. A comparison of the cumulative SFHs of our sample of simulated satellites together with observed satellites of the Milky Way and Andromeda galaxy, from the Weisz et al. (2014) sample (their Fig.7), is offered in Fig. 2: the agreement between simulations, in grey, and observations, in red, provides further evidence that our simulation provides a realistic representation of the Local Group.

As in observations (e.g. Dolphin et al. 2005), we notice in Fig. 1 that more massive galaxies show extended star formation histories, while less massive systems formed most of their stars in the first 2-3 Gyrs. A sign of the combined effect of reionization and infall quenching was found already in Rocha et al. (2012) and Benítez-Llambay et al. (2015): while classical dwarfs appear to be suppressed after infall, ultra-faint dwarfs tend to quench, for the most part, long before infall. While we do not have the resolution to study ultra-faint dwarfs, the trend between more extended SFHs and increasing galaxy mass is reproduced in our simulations: the least massive satellites only form stars in the first few Gyrs of their evolution, and by the time of their first infall, their SF has already been suppressed (see for example satellite ID=1055 or ID=310, with $\mathrm{M_{star}}\sim 10^6$ $\mathrm{M}_\odot$), while more luminous satellites whose SF is shut down, show a sign of infall quenching (see for instance ID=387 or 270, with $\mathrm{M_{star}}\sim 10^8$ $\mathrm{M}_\odot$). There is, however, an interesting finding that emerges from this figure: several satellite galaxies keep forming a considerable amount of stars after infall, some of them showing clear peaks of SF just after infall (see galaxy ID=16 or 17).

[1] www.clues-project.org
[2] www.cosmosim.org
[3] We note that there are more recent, improved SPH formulations (e.g. Read et al. 2010; Hopkins 2013), but their advantages in simulations that include star formation are not clear because of the importance of the subgrid physics modules that model star formation and feedback (e.g. Sembolini et al. 2016).
[4] http://popia.ft.uam.es/AHF

[5] In order to obtain the $AHF - ID$ used in the www.cosmosim.org database one must add 'snapnum $= 496 \times 10^6$ ' to the ID given here.

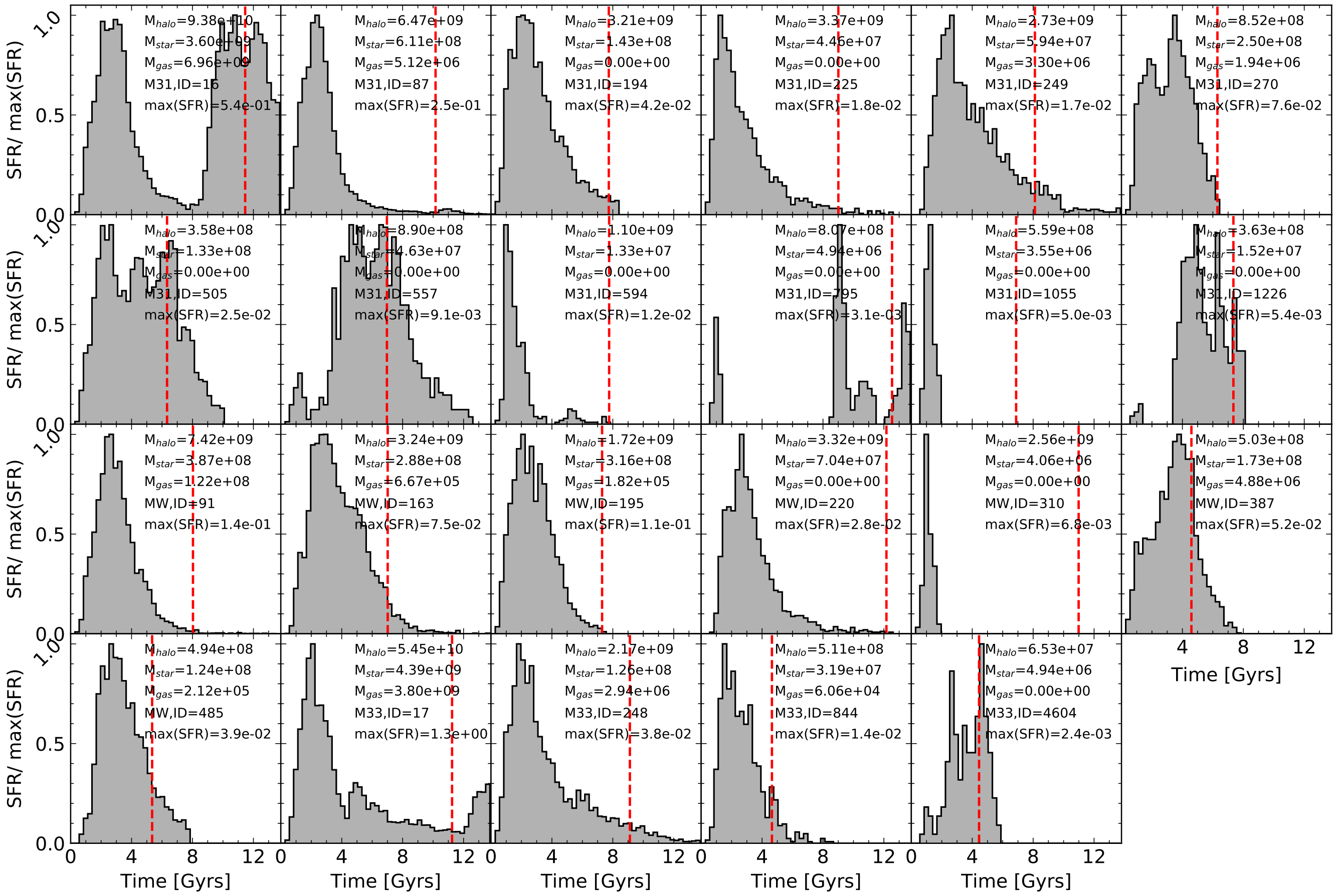

**Figure 1.** The star formation histories of satellite galaxies of M31, MW and M33, whose infall mass is larger than $10^9$ $\mathrm{M}_\odot$ and with at least 100 stellar particles at z=0. The (first) infall time is indicated as a red dashed line. The z=0 total mass of dark matter, gas and stars is shown for each satellite (measured in $\mathrm{M}_\odot$), together with the satellites' IDs. SFHs are normalized to their maximum value, indicated in units of $\mathrm{M}_\odot$ /yr. A large diversity in SFH shapes is seen, just like what found in the observed Local Group (e.g. Dolphin et al. 2005; Gallart et al. 2015, and references therein).

As such, Fig. 1 already shows one of the main conclusions of this study: *galaxies do not necessarily shut-down their star formation once becoming satellites*. We investigate and quantify this in more detail in what follows.

## 3.2 Comparison to Isolated Galaxies

We proceed next to derive the star formation histories of isolated galaxies in the simulated Local Group volume, i.e. of haloes containing at least 100 stellar particles and with $M_{\rm vir}>3\times10^8\mathrm{M}_\odot$ at redshift 0. This is a mass-selection criterion similar to that used for satellite galaxies, and takes into account that satellites will lose some of their mass since infall: it yields a sample of 83 objects. Before carrying on with our study, we remind the reader that the simulated isolated galaxies used here have already been extensively studied and analyzed in (Benítez-Llambay et al. 2015) who, in particular, showed that the SFHs of these objects resemble in many respects those of observed Local Group isolated galaxies (their Fig. 4,5 and discussion in Sec.3). We illustrate this in Fig. 3, where we show some examples of the SFHs obtained in our simulations, in several bins of mass. From top to bottom, we show the cumulative star formation of selected isolated galaxies with $10^6<\mathrm{M_{star}/M_\odot}<10^7$, $10^7<\mathrm{M_{star}/M_\odot}<10^8$ and $10^8<\mathrm{M_{star}/M_\odot}<10^9$: the left column represents a typical case of galaxies with intermediate and recent star formation, similar to what is observed in isolated dwarf irregulars, while in the right colum more ancient SF are observed, resembling the SF of dwarf spheroidals. We are reassured, therefore, that we have a variety of SF shapes resembling Local Group observations of nearby dwarfs.

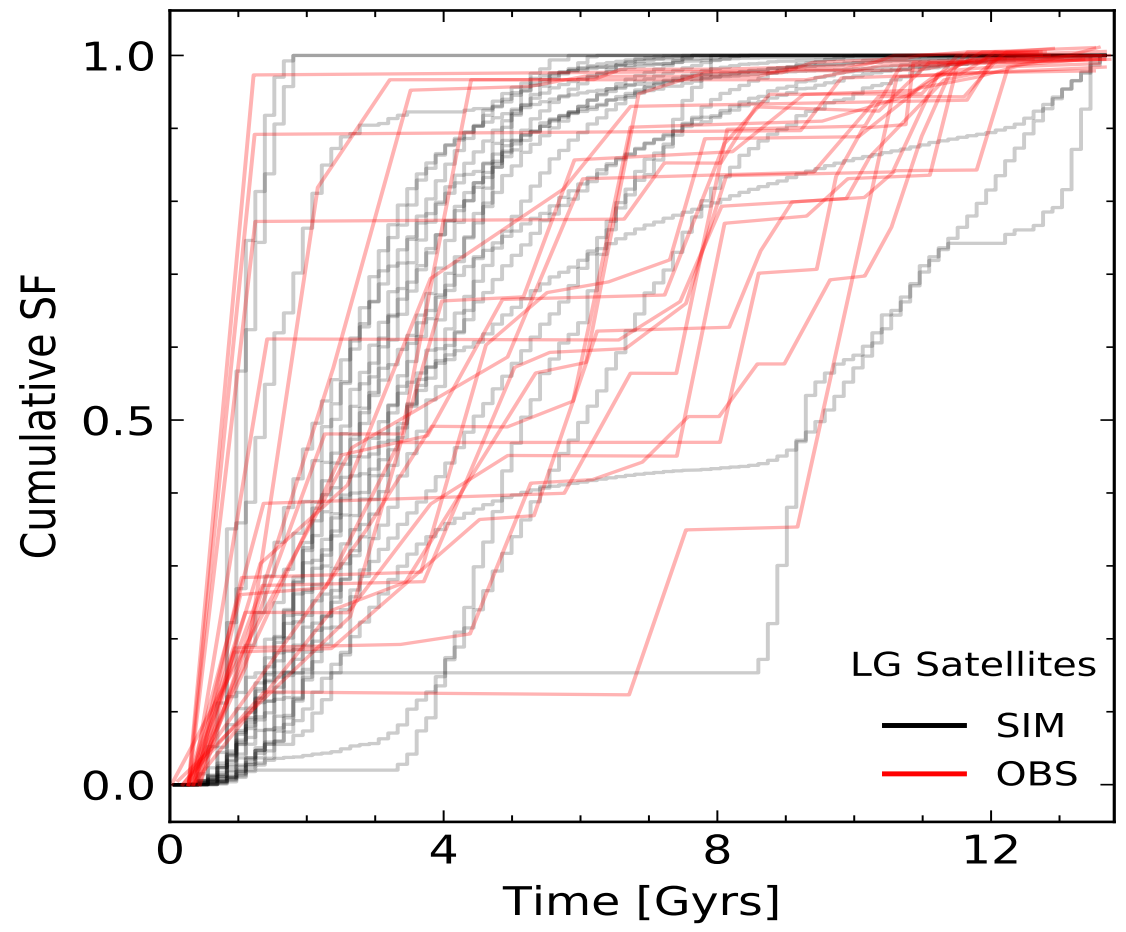

**Figure 2.** The cumulative SFH of our simulated satellite galaxies, in grey, and observed Local Group confirmed satellites, in red, including classical dSphs and dE from the Weisz et al. (2014) sample (their Fig.7). Simulated and observed satellites show similarities in their diversity of SFHs: in both groups we find examples of very old as well as intermediate-to-young stellar populations, making CLUES an ideal testbed to investigate star formation in satellite galaxies.

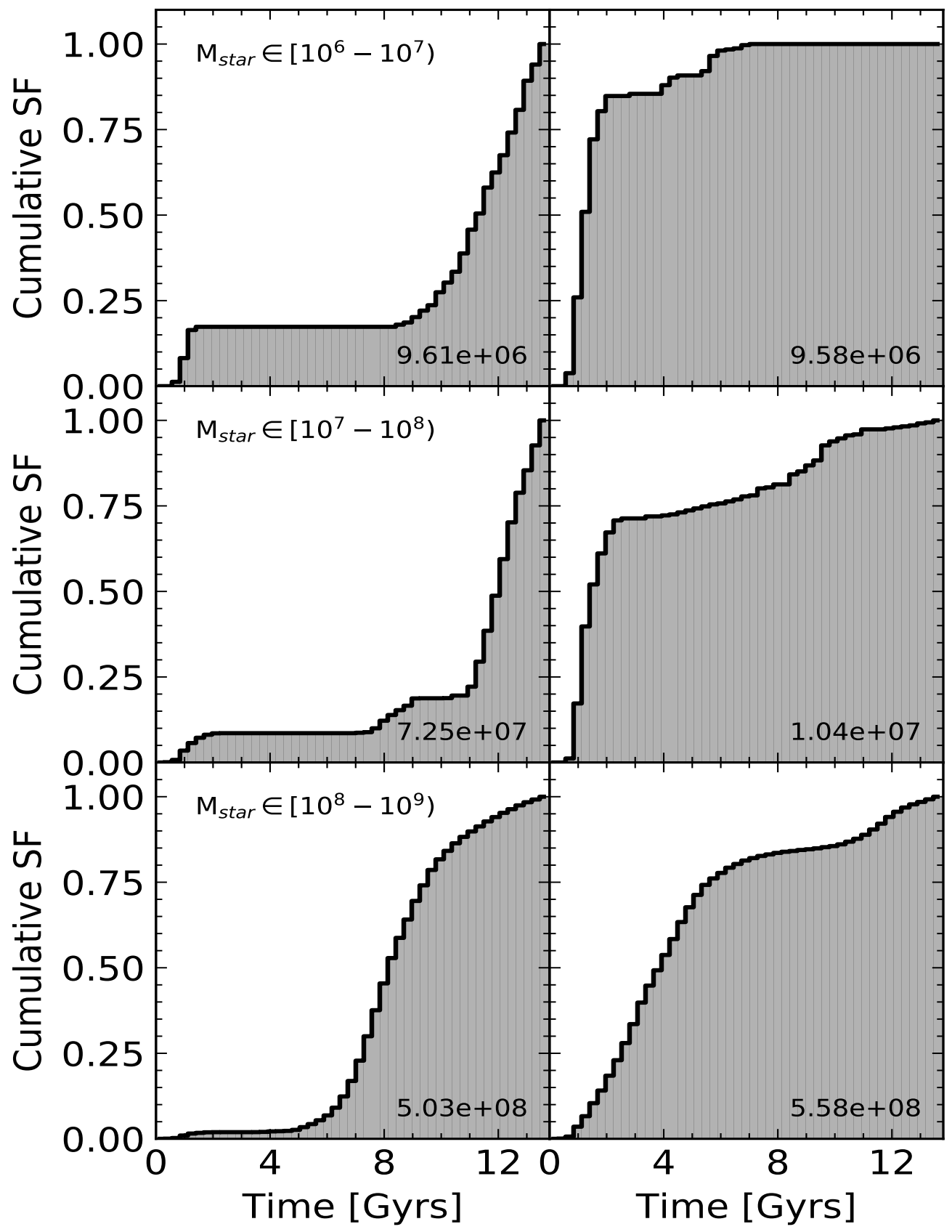

**Figure 3.** The cumulative SFH of some example isolated galaxies in our simulations: from top to bottom, we showed increasing bins in stellar mass. Galaxies in the left panel represent objects with a intermediate and young SF, similar to dwarf irregulars, while in the right column we can see more ancient SFHs, just like what is observed in Local Group's dwarf spheroidals

.

The question that we would like to address is *what is the SFH that a satellite would have had, had it evolved in isolation rather than becoming a satellite?* In order to faithfully compare the star formation history of satellite and isolated galaxies, we have to define a way to match the two samples. Associating satellites and central galaxies based on their mass at z=0 is not a good criterion, since satellites undergo substantial mass loss due to tidal stripping. For similar reasons, using the stellar mass at z=0 is not an option, as the final stellar mass in satellite galaxies will be the result of the complex interplay between star formation and infall, which is exactly the target of our study. A more reasonable and consistent approach is to assign isolated galaxies to satellites based upon the satellites' mass at the time of infall. In Fig. 4 we show as solid lines the evolution of the halo mass as a function of time for all the isolated galaxies of our sample, with the colour code scheme corresponding to progressively more massive galaxies as we move from yellow to dark purple. By inspecting the mass accretion histoy of these isolated galaxies, we noted that some of them had lost a considerable amount of mass in the last few Gyrs, a sign of the fact that they might have been not truly isolated objects in the past. We therefore applied a further cut and removed all galaxies that had lost more than 5% of their total mass in the last 6 Gyrs: this affects five objects in total, that were removed from all subsequent analysis. Superimposed in the same figure are the halo masses for the 23 satellite galaxies at infall time, each of them indicated as a red box. To a given satellite galaxy we then associate those isolated galaxies whose mass accretion history passes through the satellite's infall mass value, allowing for a $\pm 10\%$ variation in infall mass and using a 250 Myrs wide time interval: the empty red squares around each satellite's infall mass are indicative of this range. For satellites with halo infall mass between $10^9$ and $10^{10}\mathrm{M}_\odot$, this method results in as many as 13 isolated galaxies assigned to each satellite. The three most massive satellites, however, only have one isolated galaxy passing through their infall mass box, and in this case we extend the mass range (indicated as red bars in Figure 4) so that at least three isolated galaxies are assigned to each of these satellites. Once this assignment has been made, we derive the average star formation history of each group of associated isolated galaxies. We show this in Fig. 5, where each panel shows the average SFHs over the total number $N_{\mathrm{iso}}$ of isolated galaxies associated with a certain satellite, ordered in the same way as in Fig. 1. We note that there is a predominance of old stellar populations in isolated simulated galaxies compared to observations, perhaps as a result of the relatively mild feedback adopted, as already noted in (Benítez-Llambay et al. 2015). Other theoretical studies based on sets of hydrodynamical LG simulations found similar results: in Garrison-Kimmel et al. (2019) they observed that central dwarf galaxies evolving in LG environments tend to have their SFHs relatively similar to their satellite counterparts, meaning that they contain older stars than central galaxies around isolated MW-like galaxies. Therefore, the prevalence of galaxies with old stellar populations in our isolated sample can be seen as a hint that the galaxy pair MW-M31 might be able to regulate star formation in dwarf galaxies even beyond their virial radii, as suggested in Garrison-Kimmel et al. (2019). This represents an interesting avenue for future theoretical investigations. For the time being, we remark that this finding does not invalidate our results, since we are going to derive relative quantities, that is, fraction of stars in isolated compared to simulated satellites, with feedback acting similarly in both groups.

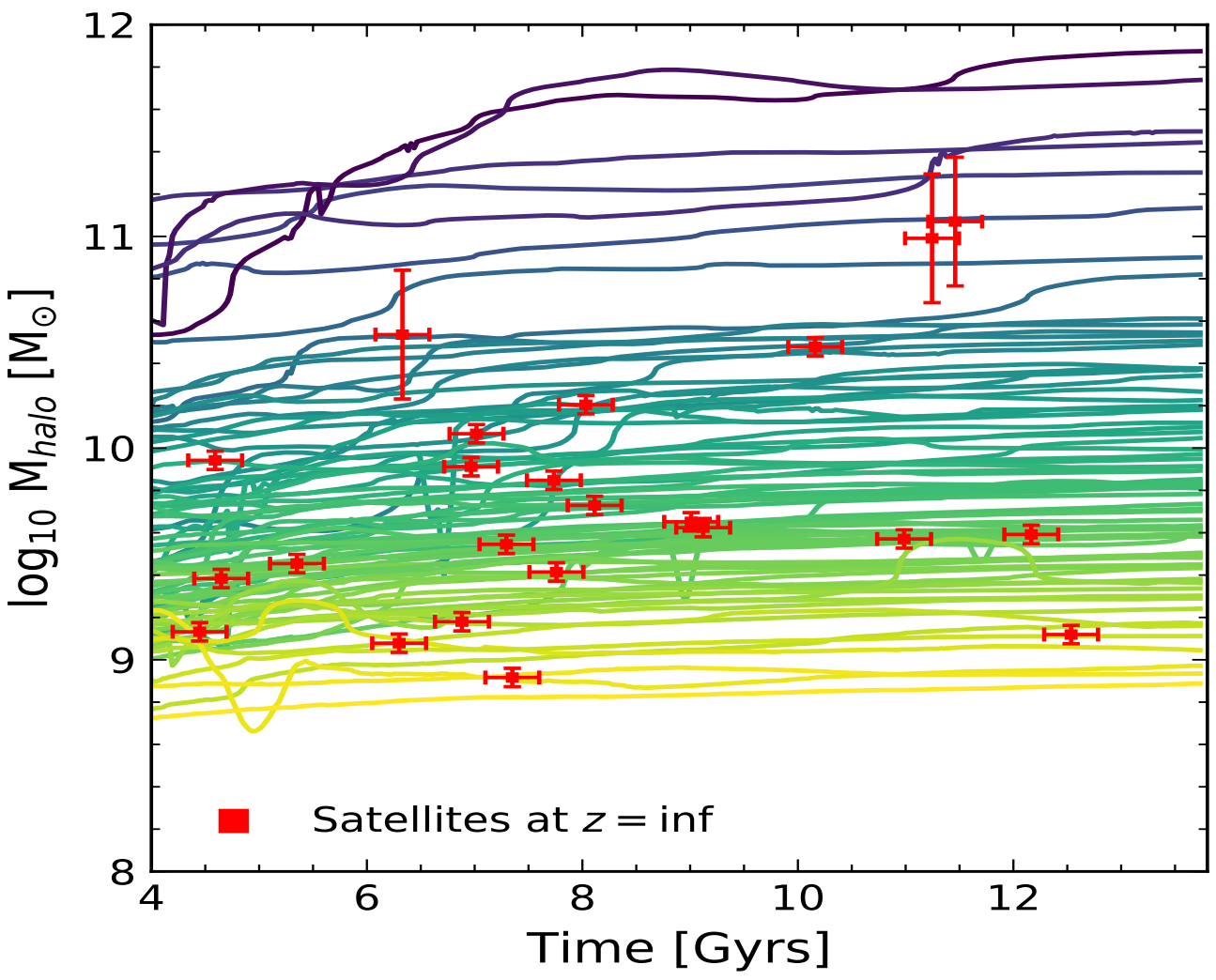

**Figure 4.** The mass accretion history of our isolated galaxies, indicated as solid colored lines (with more massive galaxies shown in purple dark, and light ones in yellow). The halo mass of each satellite galaxy at infall is indicated as a red square. We associated isolated galaxies to each satellite by matching their infall masses, allowing a $\pm 10\%$ variation in infall mass and within a 250 Myrs time interval. The three most massive satellites are connected to the three closest isolated galaxies by extending the allowed range to that indicated by the red bars, such that at least three isolated galaxies are assigned to each satellite. We have removed five galaxies that had lost more than $5\%$ of their total mass in the past 6 Gyrs, a sign that they could have been once satellites themselves.

### 3.3 Quantifying the Role of Infall

To explore what is the role of infall in shaping the SFH of satellite galaxies, we now compute for each satellite the fraction of stars formed before and after infall, $f_{\rm before}$ and $f_{\rm after}$, normalizing these fractions to the respective time interval $\triangle T_{\rm before} = T_{\rm infall}$ and $\triangle T_{\rm after} = T_{z=0} - T_{\rm infall}$. The quantity that we are interested in is:

$$\bar{f}_j = \frac{f_j}{\triangle T_j \times 2 \times \bar{f}} \qquad (1)$$

where the subscript $j$ stands for either *before* or *after* infall and

$$\bar{f} = \left(\frac{f_{\rm before}}{\triangle T_{\rm before}} + \frac{f_{\rm after}}{\triangle T_{\rm after}}\right)/2 \qquad (2)$$

such that $\bar{f}_{\rm before}$ + $\bar{f}_{\rm after}$ = 1. This procedure is similar to the one presented in Benítez-Llambay et al. (2015).

Using these definitions, care should be taken when considering satellites that might have experienced a strong stripping of stars during infall, since we are calculating the fractions $\bar{f}_{\rm after}$ and $\bar{f}_{\rm before}$ based on the SFHs obtained using the stars found within each galaxy at z=0. It is therefore possible that some galaxies experienced a decrease in their number of old stars via tidal stripping at infall time, and that this effect contributes to artificially lower the fraction $\bar{f}_{\rm before}$. This issue is particularly relevant for the satellites whose SF appears as enhanced after infall, since their respective $\bar{f}_{\rm after}$ could have been artificially increased, therefore resulting in a spurious categorization of some of these galaxies as enhanced, while in fact they are not. We take into account this effect and in what follows we re-calculate the $\bar{f}_{\rm before}$ and $\bar{f}_{\rm after}$ fractions using all the stars ever formed within a particular galaxy (i.e. the total number of stars before any stripping effect took place). This affects mostly galaxy ID=505, that has experienced a strong stripping in its stellar mass at infall, as we will see in the next section.

After computing these fractions for all our satellite galaxies as well as their associated average isolated galaxies, we can finally proceed to comparing the SFHs of isolated vs. satellite galaxies. This sheds light into the question raised earlier: what fraction of stars $\bar{f}_{\rm before}$ and $\bar{f}_{\rm after}$ would have been formed if a satellite had evolved in isolation?

## 4 RESULTS

### 4.1 Suppressed vs enhanced star formation histories

The panels of Fig. 6 show, for each satellite, the ratio between the fraction of stars formed before and after its first infall divided by the same fraction for the corresponding group of isolated galaxies. The index $i$ in the legend denotes each satellite. The ratio $\bar{f}_{\rm before,SAT_i}/\bar{f}_{\rm before,ISO}$ between the fraction of stars formed before infall in each satellite and corresponding isolated group are shown as black diamonds. The ratios $\bar{f}_{\rm after,SAT_i}/\bar{f}_{\rm after,ISO}$ between the fraction of stars formed after infall in satellites versus isolated are instead shown as circles. From top-left to bottom-right, we show such ratios as a function of z=0 stellar mass, infall virial mass, z=0 gas mass, bound gas fraction at infall, z=0 distance from host and infall time.

A ratio of one means that a satellite has formed stars just like an isolated galaxy, in the corresponding time interval. It is clear that every satellite forms stars at the same rate as their average isolated counterpart before infall, since the ratio $\bar{f}_{\rm before,SAT_i}/\bar{f}_{\rm before,ISO}$ (black diamonds) is about one for all of them. This is expected from our definition, and it is a confirmation of the fact that, before infall, there are no qualitative differences between satellite galaxies and isolated ones, in terms of their SFHs.

The situation is instead different when looking at the fraction of stars formed after infall. We define a fiducial region between $\bar{f}_{\rm SAT_i}/\bar{f}_{\rm ISO} \in 0.4dex$[6]: satellites that lie within this interval are indicated as black circles, as they behave similarly to their isolated counterparts; we shall call them 'unchanged', referring to the fact that the infall did not have any major effect on their SFH. Also unchanged are those satellites whose ratio $\bar{f}_{\rm after,SAT_i}/\bar{f}_{\rm after,ISO} = 0$ (ID=310 and 1055, again indicated as black circles), simply because their SF had already finished before infall, and therefore the infall had no impact on their evolution.

On the other hand, satellites that form fewer stars than their isolated counterparts have $\bar{f}_{\rm after,SAT_i}/\bar{f}_{\rm after,ISO} \lesssim 1$, and are indicated by blue circles in Fig. 6: for them, the infall had the effect of suppressing star formation, and we shall refer to these satellites as *'quenched'*. Oppositely, satellites that form more stars than their isolated equivalents have $\bar{f}_{\rm after,SAT_i}/\bar{f}_{\rm after,ISO} \gtrsim 1$, and are indicated by red circles; for them, the infall had instead the effect of enhancing star formation: they are the *'enhanced'* group.

Galaxy 505 is a special case, which experienced a loss of $\sim 75\%$ of its stars just after infall, and it would have therefore been spuriously categorized as enhanced if we were to use the number of stars found within it at z=0 to compute $\bar{f}_{\rm after}$ and $\bar{f}_{\rm before}$; instead, using the method explained in Section 3.3, we are sure that all galaxies, including 505, are correctly classified.

[6] Using a more strict definition of the fiducial region would not change our results, since the two galaxies (IDs= 248, 249) that have their $1 < \bar{f}_{\rm after,SAT_i}/\bar{f}_{\rm after,ISO} < 2.5$ still follow the same trends as the rest of the enhanced group.

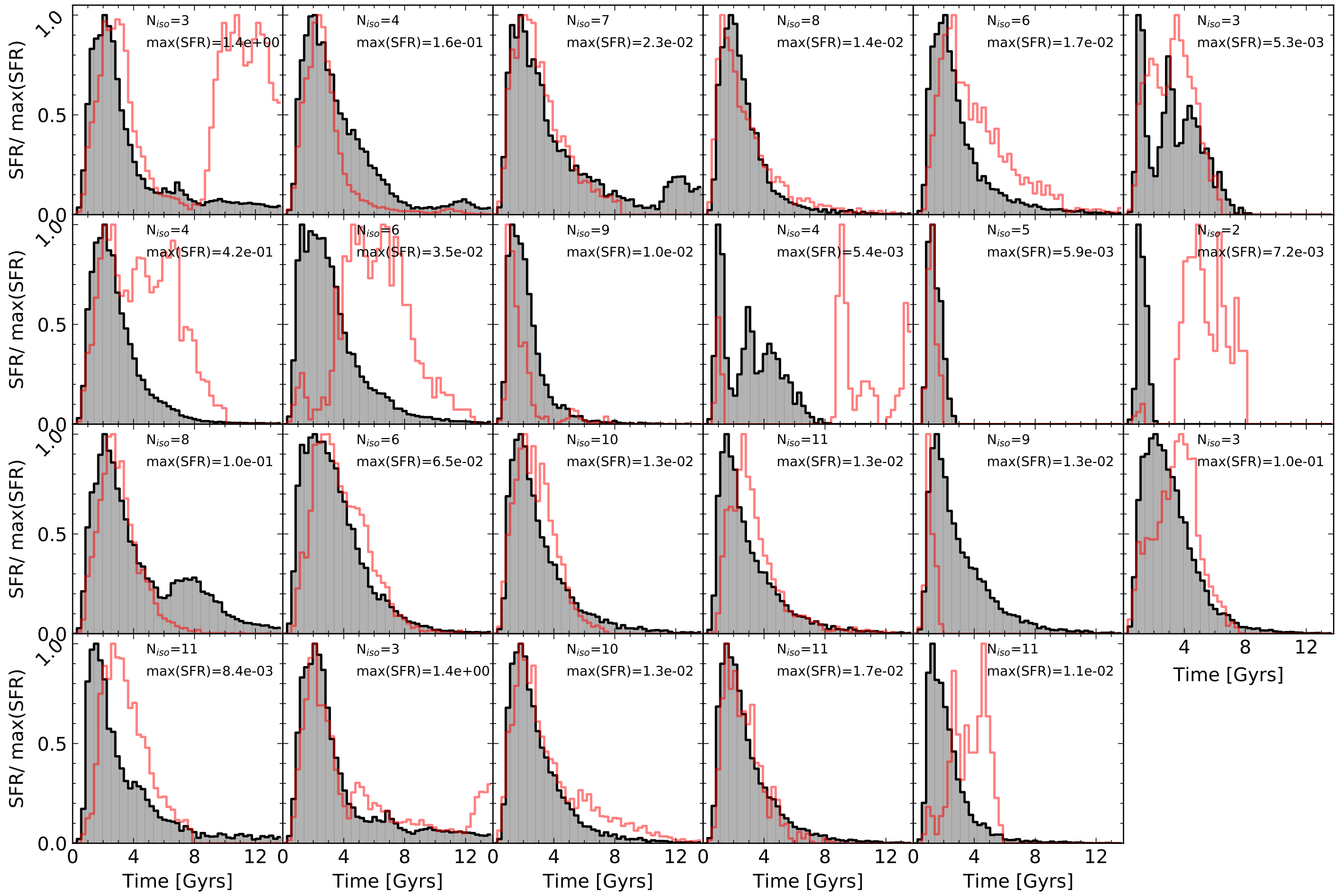

**Figure 5.** The mean star formation history of isolated galaxies whose mass accretion history matches the mass at infall of the corresponding satellite. Each box shows as a grey histogram the average over the total number (indicated as $N_{\rm iso}$) of isolated galaxies associated to a particular satellite, following the same order as Fig. 1. As a red line we indicate the SFH of each corresponding satellite.

We are left with a final sample of 6 bona-fide enhanced satellites, indicated as red circles in Fig. 6: that 6 out of 23 satellites are found to have their SFH enhanced after infall ($\sim$25%) is a surprising result, since typically galaxy formation models assume that most satellites shut down their SF after infall (e.g. Starkenburg et al. 2013, and references therein).

Fig. 6 further indicates that there is no apparent correlation between the enhancement (or the suppression) of a satellite' SFH and either the stellar or gas mass at z=0, the virial mass at infall, the final distance from host or the infall time, although for increasing stellar masses the fraction of suppressed satellites diminishes, such that in the range $10^9 < {\rm M_{star}}/{\rm M_\odot} < 10^{10}$ none of the satellites is SF suppressed, as reported in Slater & Bell (2014) (note, however, the small numbers in our sample in this regime). The quenching fractions in our simulations seem to be in agreement with previous literature based on numerical simulations (e.g. Simpson et al. 2018; Akins et al. 2021): in Simpson et al. (2018), for instance, the authors found that for stellar masses larger than $10^9$ M$_\odot$ the quenched fraction of satellites dramatically declines, while for stellar masses between $10^{8-9}$ M$_\odot$ such fraction is a strong function of the distance from the host, with up to 50% quenched satellites found for a distance 0-0.3 Mpc in their simulations. In our sample, there is no quenched satellite with ${\rm M_{star}}{>}10^9{\rm M_\odot}$, and only 4 out of 10 satellites are strongly quenched in the stellar mass bin $10^{8-9}$ M$_\odot$, while the rest are left unchanged.

Interestingly, the middle-right panel of Fig. 6 indicates a clear correlation between the fraction of bound gas at infall (defined as the gas within 20% of the virial radius of the satellite divided by the satellite's virial mass, both quantities calculated at infall) and the enhancement in star formation. We explore this finding next.

### 4.2 The role of gas and pericentric passage

To further investigate the reasons behind the different evolutions in satellite's SFH after infall, we show once more in Fig. 7, left panel, the gas fraction at infall as a function of enhancement fraction, while in the right panel we show the same quantity but as a function of minimum pericentric distance, in kpc. Satellites are again color coded in red, if their SFH is enhanced after infall, and in blue if their SFH is suppressed. Comparing these two panels provides hints about the mechanisms that cause the enhancement of SF in some galaxies but not in others.

Firstly, it is evident that the SF enhanced satellites are those objects infalling with a high fraction of cold gas, i.e. with a value of $M_{\rm gas,inf}(<20\%R_{\rm vir})/M_{\rm vir,inf} \gtrsim 10^{-2}$. We verified that all the gas particles found within the 20% of the satellites' radius at infall are either in a cold or star forming phase with temperatures of $T< 10^5$K and densities $n> 0.12cm^{-3}$. Furthermore, we have checked that none of the satellites infalling with less than $10^7 M_\odot$ in gas, in absolute value and within their central region, is able to form new stars after infall.

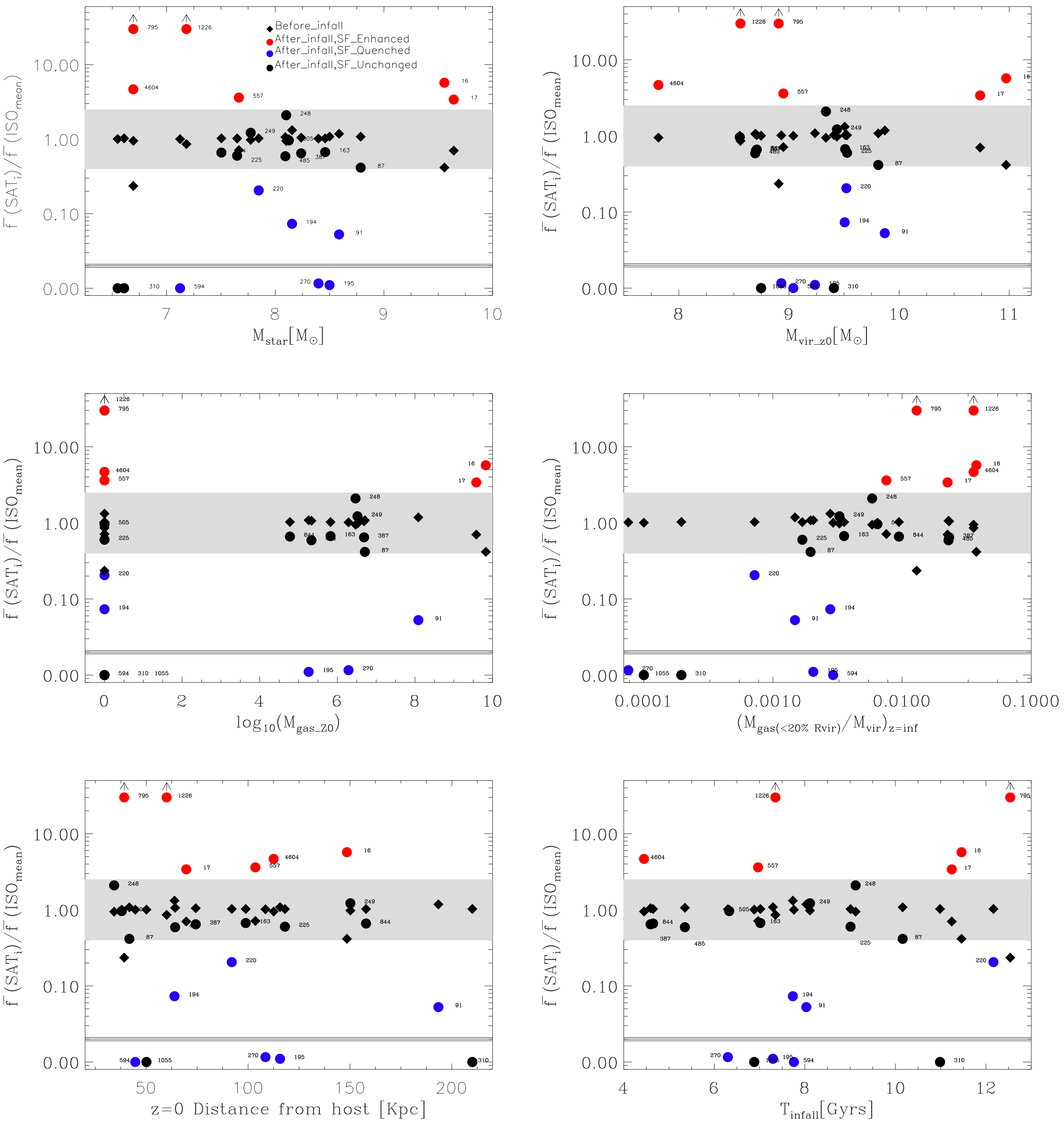

**Figure 6.** The normalized fraction of stars formed before ($\bar{f}_{\rm before}$, diamonds) and after infall ($\bar{f}_{\rm after}$, circles) for each satellite, divided by the same number for the average isolated galaxy group corresponding to each satellite. From top-left to bottom-right, we show this ratio as a function of z=0 stellar mass, total virial mass and gas mass, fraction of bound, cold gas at infall, z=0 distance from host galaxy and infall time. Satellites which have their star formation enhanced after infall are indicated as red circles, satellites whose star formation is instead suppressed are indicated as blue circles. Satellites whose SF has not changed after infall, compared to the isolated group, are indicated as black circles. Enhanced, quenched and unchanged satellites are selected taking into account a fiducial region of $\bar{f}_{\rm SAT_i}/\bar{f}_{\rm ISO} \in 0.4dex$. Black diamonds refer to stars formed before infall, which are similar in satellites and isolated galaxies. Satellite IDs are shown for comparison with Fig. 1. Satellites ID=795 and 1226 have an enhancement ratio larger than the plot limits, and are therefore indicated with an upwards arrow. Satellites ID=310 and 1055, whose ratio is $\bar{f}_{\rm after,SAT_i}/\bar{f}_{\rm after,ISO} = 0$, are categorized as unchanged and indicated as black circles, simply because their SF had already finished before infall, and therefore the infall had no impact on their subsequent evolution. Within our sample of 23 simulated satellites, as many as 6 are bona-fide enhanced galaxies, making up for $\sim$25% of the sample size. No correlation is observed between the enhancement ratio and any of the quantities shown, except in the case of gas fraction in the inner region of the satellite at infall time: enhanced satellites enter their host's virial radius with a relatively large gas fraction, $M_{\rm gas,inf}(< 20\% R_{\rm vir})/M_{\rm vir,inf} \gtrsim 10^{-2}$.

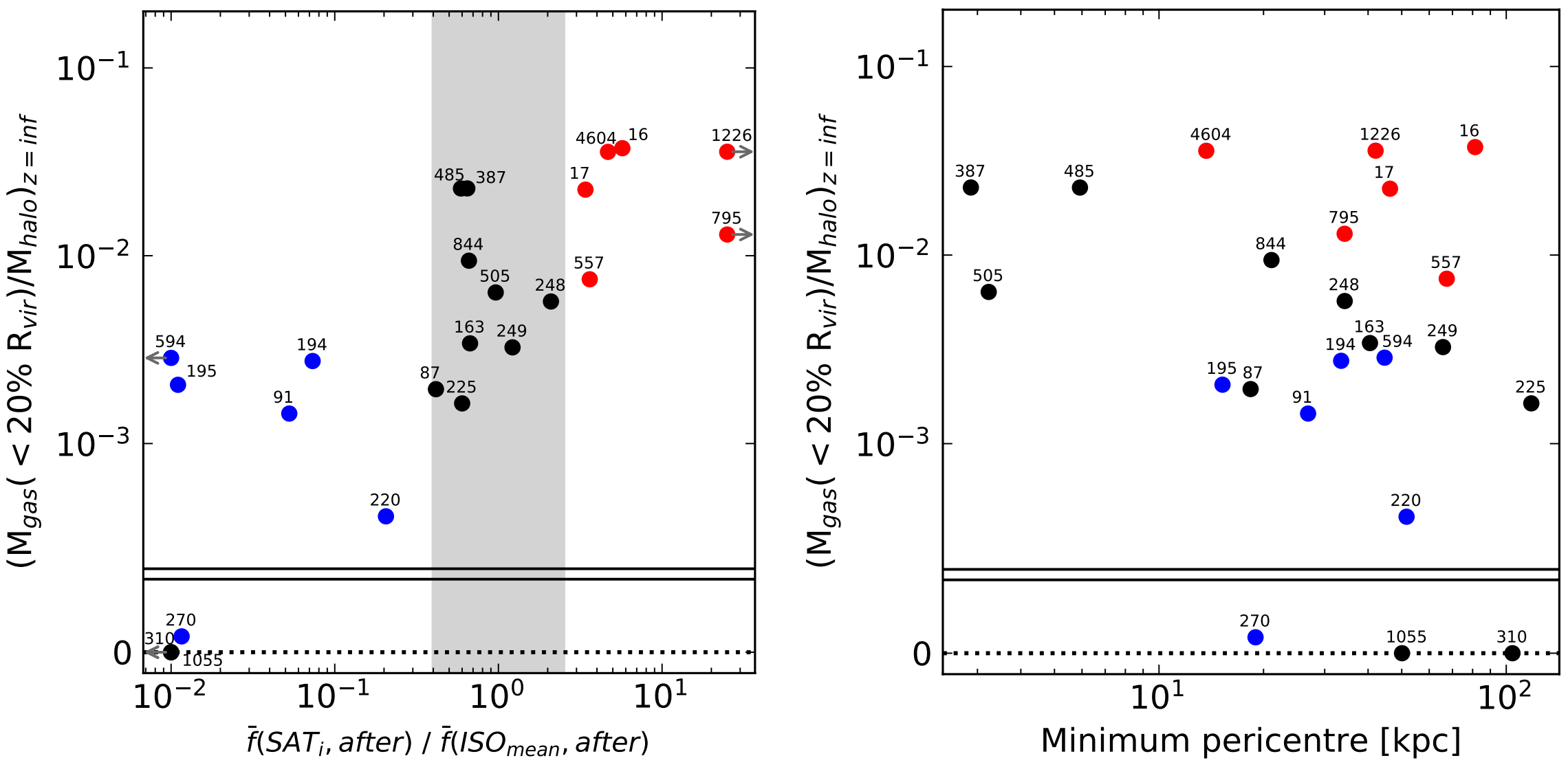

**Figure 7.** The ratio of gas at infall within 20% of the virial radius of each satellite, normalized to the satellite's virial mass, versus enhancement ratio (left panel) and minimum pericentric distance from host (right panel). The color code is the same as in Figure 6, with enhanced satellites indicated in red, quenched in blue and unchanged in black (within the fiducial region shown as grey area). Those satellites that infall into their host galaxy with a large amount of cold gas, and nevertheless do not show an enhancement in their SFH, are those whose minimum pericentre is small, thus enabling the loss of such gas via ram pressure stripping. Oppositely, if a galaxy infalls into its host with lots of bound gas and, additionally, its orbit results in a larger pericentric distance (minimum of 13 kpc, average of ∼47 kpc), the compression of such gas can result into a burst of newly forming stars.

However, there are a few satellites that, despite having large amounts of cold, star forming gas at infall, do not show any enhancement in their SF (namely IDs= 387,485): the reason for this must be searched in their orbits.

In the right panel of Fig. 7 we see that these objects are on very eccentric orbits, so that their pericentric distance is only a few kpc, resulting in a net loss of gas by ram pressure stripping. Conversely, enhanced satellites keep forming stars after infall because their pericenter distance is larger (with a minimum of 13 kpc and an average of ∼47 kpc for the whole enhanced group), which allows the existing cold gas to be compressed during pericentric passage, rather than lost due to stripping: it is the compression of such gas at pericentric radius that leads to new bursts of SF in the enhanced satellites group. Quenched satellite galaxies, on the other hand, do not form stars after infall, regardless of their orbit, since they did not retain enough gas in the first place.

So far, our results suggest that *in order for a satellite to enhance its SF after infall, it must enter into its host galaxy with large reservoirs of cold gas and on an orbit whose pericentric distance is not too small*. We now consider examples of enhanced vs unchanged satellites, and additionally show that new stars form in enhanced satellites exactly at the time of pericentric passage around their hosts.

In Fig. 8 we show the three most relevant examples of enhanced satellites, from left to right we show ID=16, 17 and 795. In the top panels we indicate in solid red the virial radius of their host galaxy, and in black the orbit followed by the satellite, both quantities as a function of cosmic time. The point at which the two lines cross each other corresponds to the infall time, also indicated as a dashed vertical red line. In the case of ID=795, we further indicate, with a black dashed line, the orbit of this satellite with respect to another dwarf galaxy, in this case ID=16. In the bottom panels we show the normalized SFHs of such enhanced galaxies, together with their properties at z=0. It is evident the strong correlation that exists between the pericentric approach to the host galaxy and a burst of newly formed stars in the satellites, clearly visible in all of the studied cases, and happening respectively at ∼12, ∼13 and ∼12.5 Gyrs. Moreover, in the case of satellite ID=795, a large burst of SF is also seen at ∼9 Gyrs, correlating with the close-by passage with satellite ID=16 (the closest approach is reached at 8.87 Gyrs and at a distance of 48.5 kpc), suggesting that this satellite-induced enhancement could also happen in the real Universe. At pericentric passage, either around their host or around another dwarf, as in the case of ID=795, the total amount of gas decreases only slighty in these satellites, accounting mainly for the conversion of gas into new stars.

In Fig. 9 we show similar results but for those satellites that infall into the main host with plenty of cold gas but nevertheless do not show enhancement in their SF, relative to the isolated galaxies group. The reason is again related to their orbits: they experienced an extremely close-by pericentric passage, which is able to ram pressure strip most of their gas and dark matter. These satellites lose about 99.9% of their gas just after the pericenter passage, going from a total gas mass of $\sim 10^8 M_\odot$ to $\sim 10^4 M_\odot$ in less than 3 Gyrs. Some residual SF is observed during this time, which is then completely quenched once most of the star forming, cold gas, has been stripped away. These objects survive despite their extreme orbits; their z=0 properties, however, are a faithful reflection of their violent past: most of their dark matter has been stripped away from the outskirts, such that the leftover dark matter mass is only 4-5 times the mass in stars, that, being centrally concentrated, undergo less stripping (less than 5%). Our galaxies are resolved enough and compact enough to ensure that a satellite survives under these severe conditions, and we refer the reader to the discussion in van den Bosch & Ogiya 2018 and Errani & Peñarrubia 2020 about numerical disruption of satellites in lower resolution regions and, more in general, to Buck et al. 2019 and references therein for analysis of how mass loss in orbiting satellites directly correlates with the minimum distance ever reached.

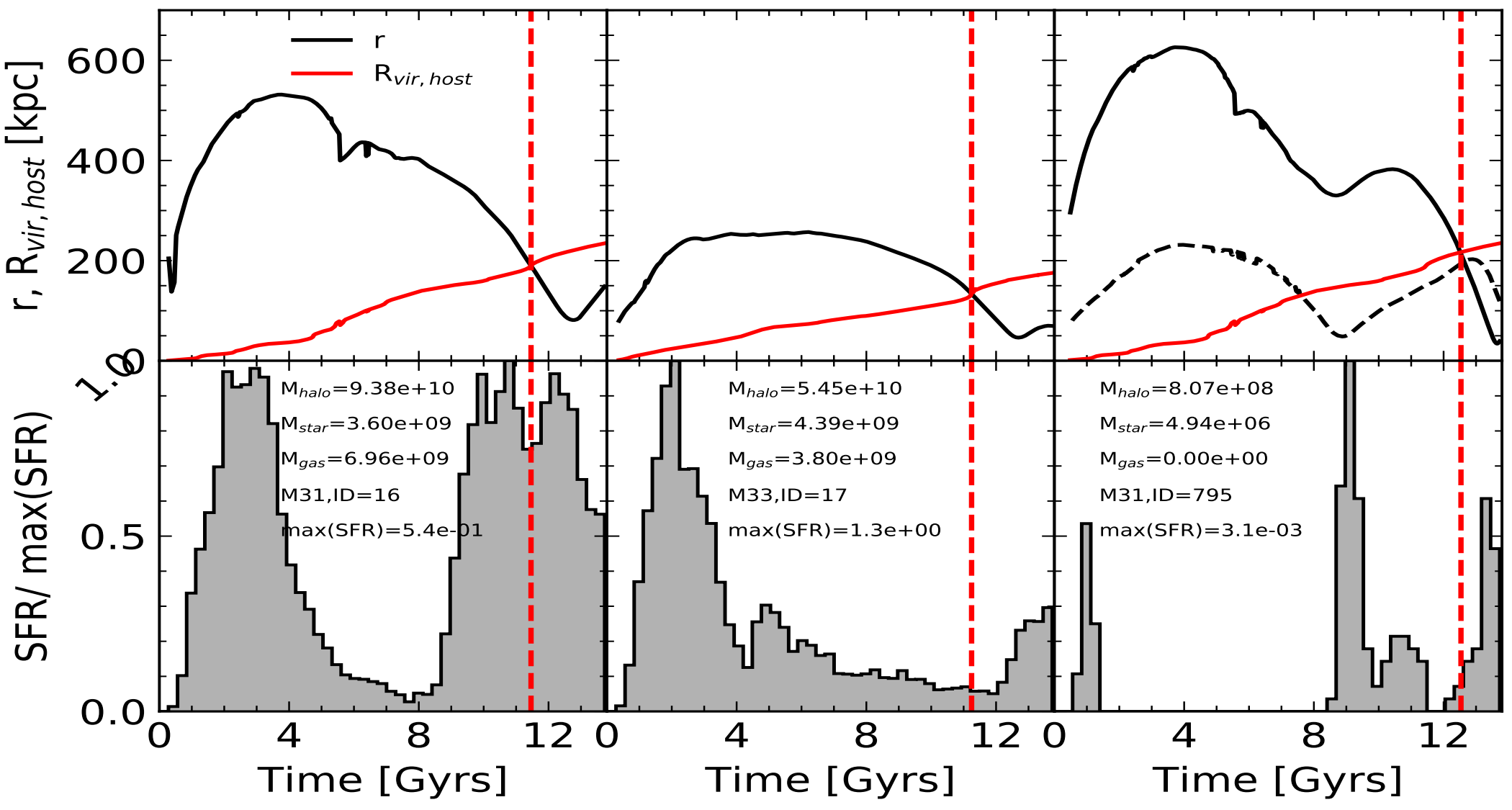

**Figure 8.** Examples of star formation-enhanced satellites: from left to right, we show satellites ID=16, 17 and 795. For each satellite we plot, as a function of time and from top to bottom, its orbit with respect to the main galaxy in black, the main host galaxy virial radius in red, and the satellite' SFH normalized to its maximum value. The vertical dashed line corresponds to the infall time within the host galaxy. For satellite=795 we also show its orbit with respect to another dwarf galaxy, namely ID=16, as dashed black line. It can be observed that in all cases star formation is enhanced exactly at the time of pericentric passage, due to the interaction between the satellite and its host, and the resulting compression of existing cold gas which in turn forms new stars. This is achieved thanks to the fact that the pericenter distance of these satellite-host systems is not too small, being, respectively, 83, 47 and 35 kpc for satellites ID=16,17 and 795. In the case of satellite ID=795, a dwarf-dwarf close-by interaction, with a minimum distance of 48.5 Kpc, is further shown to be responsible for an increase of SF at around ∼9 Gyrs. In our simulations, the only galaxies that undergo a burst of SF at pericenter passage, after infall, are the ones falling into the 'enhanced' category: while for the 'unchanged' group there might still be some leftover SF happening when the satellite passes at pericenter, such SF is not influenced by the pericenter itself, i.e. no burst is observed.

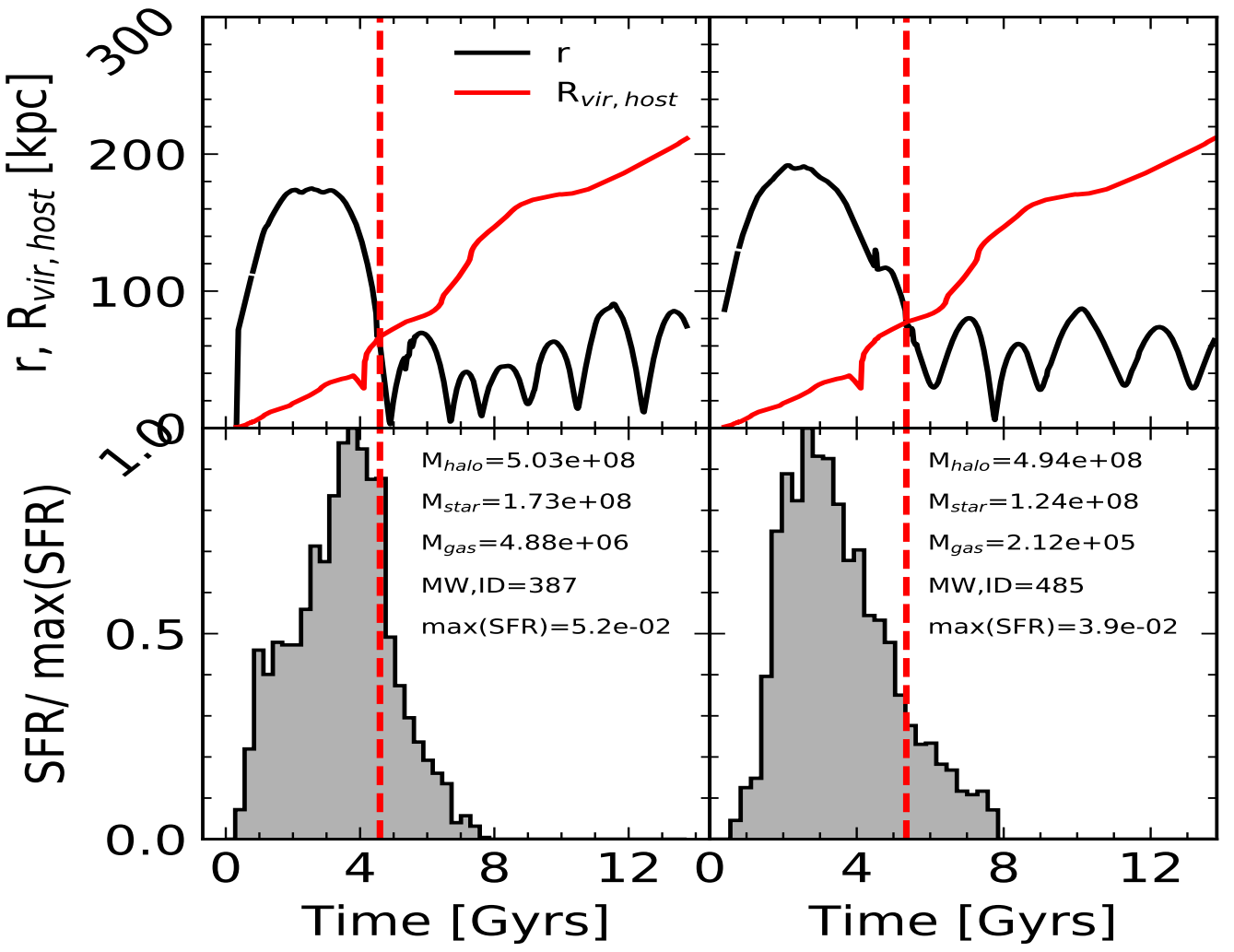

**Figure 9.** Same as Fig. 8, but this time for those satellites that, despite infalling into their main hosts with plenty of cold gas, $M_{\rm gas,inf}(< 20\%R_{\rm vir})/M_{\rm vir,inf} \gtrsim 10^{-2}$, do not keep forming stars (namely ID=387, left panels, and ID=485, right ones). Here, the star formation is gradually quenched due to the severe gas loss provoked by the extremely close-by approach of these satellites with their hosts (pericenter of 2.8 and 5.8 kpc).

To conclude this subsection, we show in Fig. 10 the gas density distribution at different subsequent redshifts for two SF-enhanced satellites (ID=17 and 795, first two columns) and one SF-unchanged (ID=387, last column). These gas density maps have been derived using the publicly available *Py-SPH-viewer* code [7]. The color code, from red to blue, indicates regions of high to low gas density. The high gas density regions correspond to star forming areas. Each panel is centered on the satellite galaxy studied, and the black circle represents the satellite's virial radius. We show different temporal snapshots corresponding to the time of infall of each satellite around its main host.

In the first column we study the evolution of satellite ID=17: this satellite follows the average SFH of relatively massive galaxies until infalling into its host at 11.5 Gyrs; at this point, the interaction with its host galaxy, clearly visible in the gas density time-snapshots, causes a new peak of star formation just after 12 Gyrs. Please note that satellite ID=17 is one of the most massive satellites in our sample, and has indeed a very extended star forming region, such that it looks similar to its main host, M33, in these images (although of course being less massive than M33). In the middle column we look at satellite ID=795: this galaxy, being very small at reionization, only formed a few stars before z=6, then it evolves without forming any stars until a cosmic time of 8 Gyrs. During this period, subsequent mergers visible in its mass accretion his-tory (not shown here), allow this galaxy to grow in both dark matter and gas; the close encounter with another dwarf satellite galaxy, namely ID=16, visible at around 8.87 Gyrs (first 3 panels of the central column) will compress the gas and give rise to a second SF episode, while the final approach towards the main host halo M31, at 12 Gyrs, will provoke a final peak in SFH (last two panels). In the last two panels, the two large galaxies shown are satellite ID=16 and the main host M31: at the depicted time, these two galaxies are found at a distance of about ∼83 kpc from each other, being ID=16 just reaching its minimum pericenter around M31.

[7] https://github.com/alejandrobll/py-sphviewer

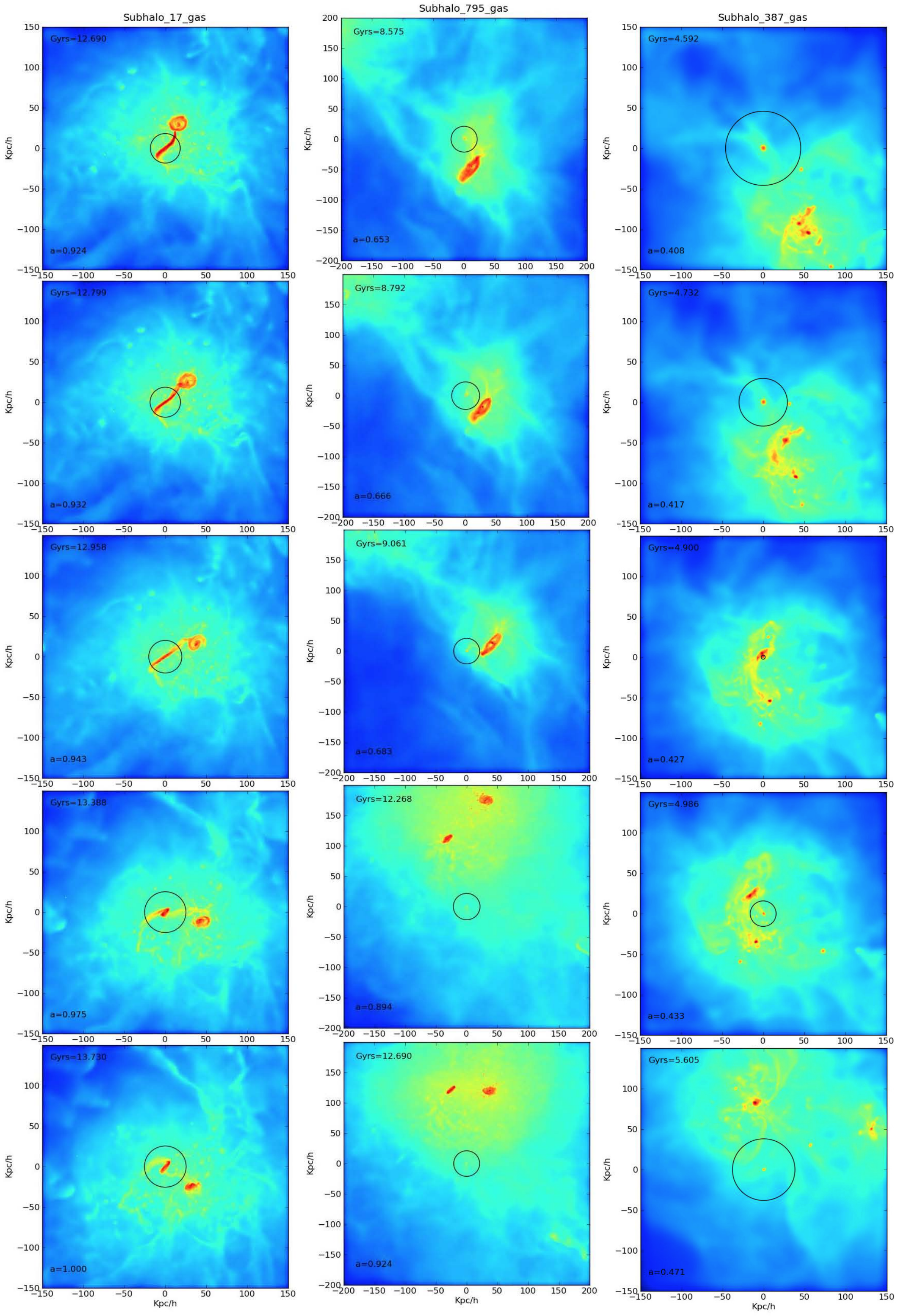

**Figure 10.** Gas density maps at subsequent snapshots around infall redshift for satellites 17, 795 (enhanced SF) and 387 (unchanged SF). The evolution of the gas distribution is shown at increasing times from top to bottom. Each panel is centered on the satellite galaxy analyzed, with its virial radius indicated as a black circle. The color scheme indicates high (in red) and low (in blue) gas densities. In the first column, the interaction of the satellite with its host galaxy can be appreciated: the pericentric passage in this case is $\sim$47 kpc, and the net result is an increase of the satellite SF. In the middle column, a close approach between the satellite galaxy and another dwarf can be observed in the first 3 panels, happening at around 8.5 Gyrs and provoking an episode of star formation, with a subsequent interaction with the main host at $\sim$12 Gyrs causing the final SF burst (the pericentric distance in this case is about 35 kpc). Finally, in the last column, an extremely close-by approach, with minimum distance of 2.8 kpc, is seen, which removes most of the existing gas from the infalling satellite: its SFH is affected such that it completely stops 3 Gyrs after infall has occurred. Movies of the cosmic evolution of these satellites are found at `https://dicintioarianna.wixsite.com/mysite/movies`.

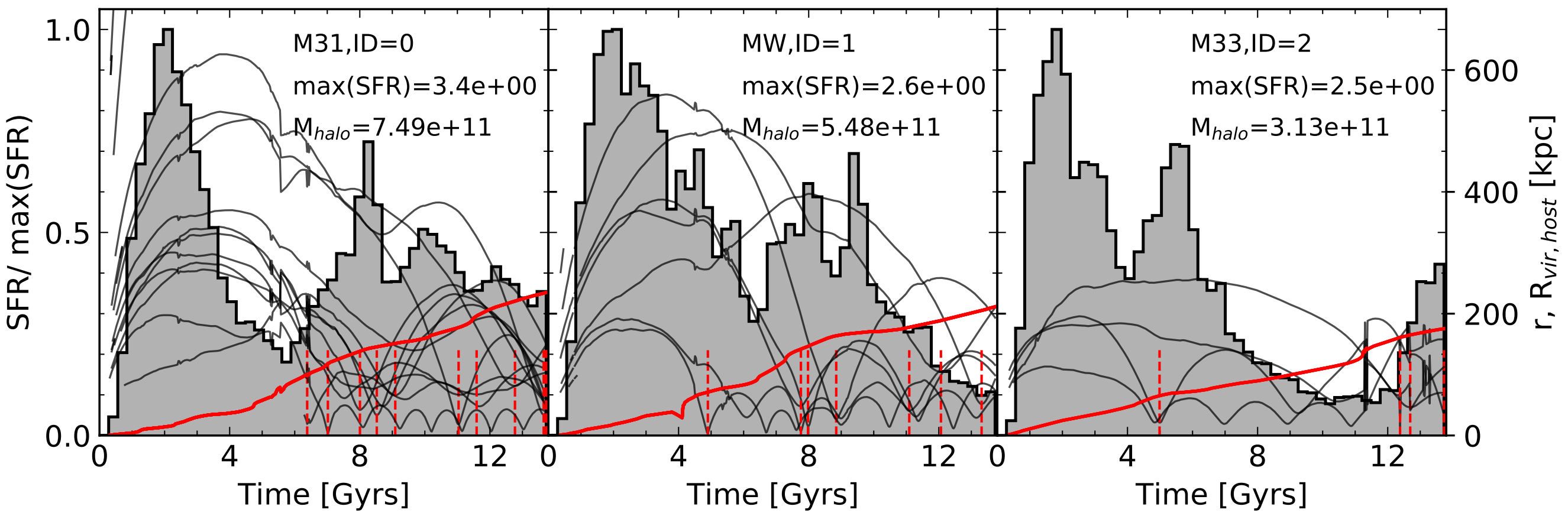

**Figure 11.** The SFH of the three main galaxies in our simulation, namely M31, MW and M33, with their virial radius indicated as a solid red line. The maximum SFH is indicated in each panel, in units of $M_\odot$ /yr. Overimposed as thin black lines are the orbits of their respective infalling satellite galaxies: we indicated as vertical dashed red line the time of minimum pericentric approach of each satellite. Clear peaks of SF can be identified in the host galaxy at the time when a satellite reaches its pericenter, with this effect being more pronounced when multiple satellites approach the host at similar time, a sign of the impact of 'group infall' on the main galaxy SFH. This is reminiscent of the recently reported impact of the Sagittarius dwarf on the Milky Way SFH (Ruiz-Lara et al. 2020).

The SFH of galaxy ID=795 is particularly interesting since it is reminiscent of the episodic formation history of Carina dSph, whose SF also shows episodic peaks at old and intermediate ages, and whose structural differences argue against an isolated evolution of this galaxy (de Boer et al. 2014). Therefore, an intriguing explanation for the SFH of Carina dwarf galaxy, could be a past close interaction with other dwarf systems, before its final infall into the Milky Way's potential. Finally, in the right-most and last column we indicated the evolution of satellite ID=387, whose close interaction with its host causes a dramatic reduction of its gas content just after the interaction. Indeed, a sharp decline in SF is visible exactly at 4.5-5 Gyrs. In isolation, this galaxy would have continued forming stars all the way until exhausting its reservoir of cold gas, and it would have been characterized by having both an old as well as a young stellar population: however, due to the effects of an extremely small pericentric passage, its reservoir of gas is depleted via ram pressure stripping, and the galaxy effectively stops forming stars within 2-3 Gyrs from infall.

### 4.3 Impact of pericentric passage on the host's SFH

In this section we show that satellite-host interactions not only affect the subsequent SFH of satellites, but that they may have a tangible impact on the SFH of the host galaxy itself. In Fig. 11 we show the normalized SFH of the three main host galaxies in our simulation, namely M31, MW and M33, with their virial radiii indicated as a solid red line across cosmic time. The maximum SFH is indicated in each panel, in units of $M_\odot$/yr. Overimposed as thin black lines we show the orbits of their satellite galaxies, the same ones analyzed so far. We further indicate as vertical dashed red lines the time of minimum pericentric approach of each satellite.

The SFH of the three host galaxies show clear signs of the impact of their infalling satellites: indeed, peaks of SF can be identified at the time at which the satellites approach pericenter, with this effect being more pronounced when multiple satellites approach the host at a similar time. This is in line with the newly reported finding of the impact of Sagittarius dwarf on the Milky Way' SFH (Ruiz-Lara et al. 2020): Gaia DR2 data have revealed three episodes of enhanced star formation in the Milky Way, whose timing coincides with the proposed pericenter passages of the Sagittarius dwarf.

While a more in-depth study of such effect would be desirable in the future, it is clear from these results that perturbations induced by an infalling satellite, or by several of them, are able to trigger episodes of star formation in their main host galaxy.

### 4.4 Comparison with Gaia observations

We conclude the result section by showing, in Fig. 12, a comparison of our simulated satellites with observed Local Group dwarf galaxies within 420 kpc of the Milky Way, whose orbital properties are collected in Fritz et al. (2018) using Gaia DR2 data. Specifically, in Fritz et al. (2018), the values of apocenter, pericenter and eccentricity of each dwarf are derived by using the publicly available code *galpy* to integrate the dwarf's orbits in a Galactic potential, assuming a Milky Way virial mass of $1.6\times10^{12}M_\odot$, in line with recent estimates (see Fritz et al. 2020 and references therein).

In the left panel we show the radial velocity versus pericenter, while in the right panel the eccentricity versus radial velocity is indicated. For the simulated satellites we used the closest pericentres and apocentres to z=0, so that we could safely compare them to observational data, and we computed the eccentricity as $ecc = (r_{apo} - r_{peri})/(r_{apo} + r_{peri})$. In some cases the satellite is just infalling into its main host (e.g. ID=795), or has not reached the apocenter yet (e.g. ID=16) therefore the estimation of the eccentricity is only approximate in these cases.

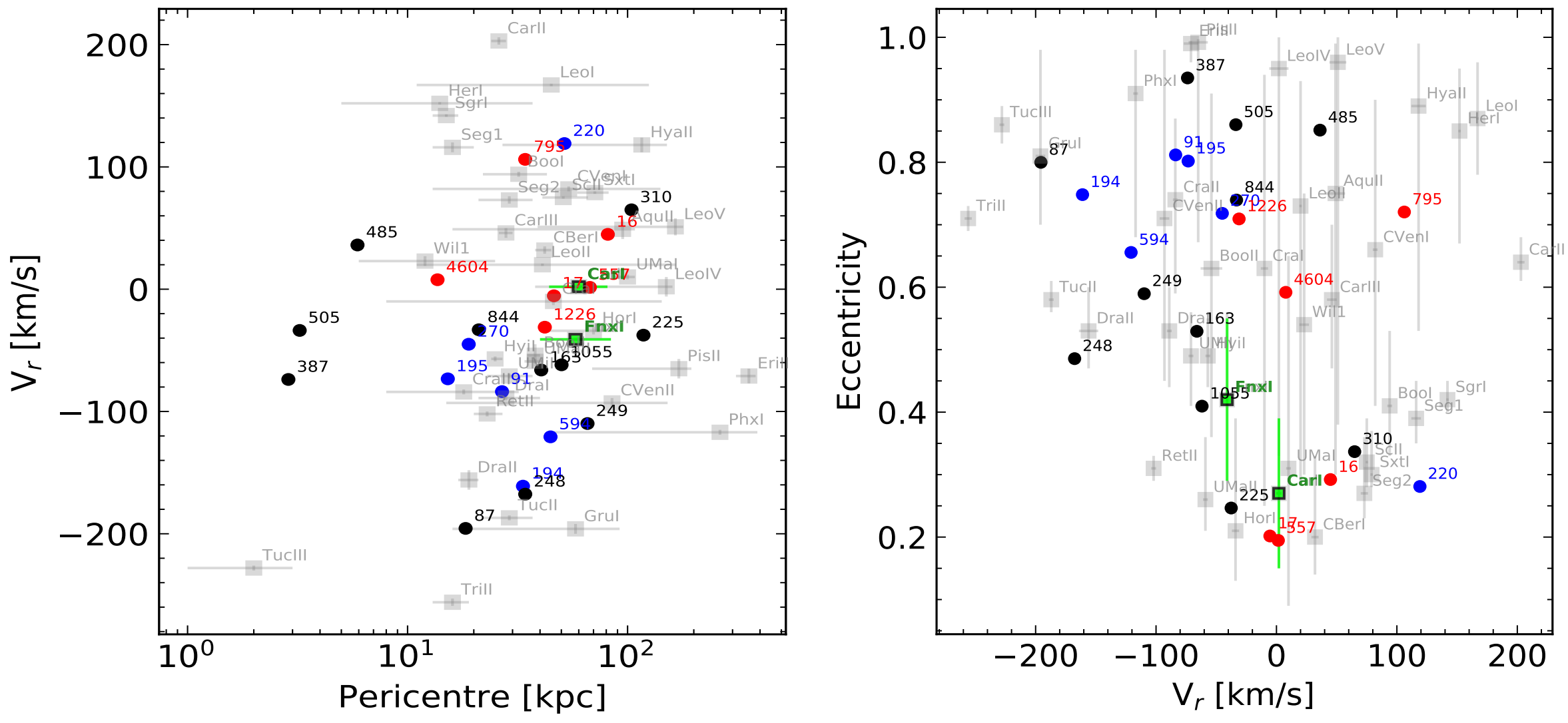

**Figure 12.** Comparison between CLUES simulated satellite galaxies (circles) and observed Local Group dwarfs (grey squares with error bars) from Gaia DR2 data (Fritz et al. 2018). Simulated enhanced satellites (in red) have a positive or zero radial velocity and lie on nearly circular orbits (3 out of 6 cases), while quenched/unchanged satellites (blue and black circles) have on average a much higher value of their eccentricity parameter, computed as $ecc = (r_{apo} - r_{peri})/(r_{apo} + r_{peri})$. Fornax and Carina galaxies (highlighted in green), showing episodic star formation possibly related to pericentric passages around the Milky Way (Rusakov et al. 2021), have a similar $ecc$, $V_r$ and pericentre value as our enhanced group, supporting the scenario in which satellites whose pericentric distance is not too small, and additionally infall with large gas reservoir, might suffer a burst of SF triggered by such interaction.

Nevertheless, we can appreciate that the simulated satellites and observed LG dwarfs populate a similar phase space in the $V_r - peri$ and $ecc - V_r$ planes. Enhanced satellites, again indicated as red circles, tend to have a radial velocity close to zero (4 out of 6 cases) or positive (remaining 2 satellites), indicating they are currently moving away from their host, or are orbiting around it at pericenter/apocenter. As opposite, most of the quenched and unchanged satellites have a negative value of $V_r$. The average eccentricity of the enhanced group is clearly smaller than the one of the quenched and unchanged group, with three satellites being on nearly circular orbits (ID=16,17 and 557). It is intriguing to note how Fornax and Carina galaxies, the two satellites that show clear signs of multiple stellar populations whose connection with the pericentric passage has been argued (e.g. Rusakov et al. 2021) also have a relatively large pericentric distance, of about 58 and 60 kpc, respectively, and lie on rather circular orbits (eccentricity of 0.42 and 0.27), just like several of our simulated enhanced satellite galaxies. We stress that a more in depth interpretation of these results will require further work, given the current large observational errors and the small number statistics of our simulated sample.

## 5 CONCLUSIONS

We used cosmological simulations of the Local Group of galaxies, from the CLUES project (Gottlöber et al. 2010; Carlesi et al. 2016), to study the star formation histories (SFH) of the satellite galaxies of the three most massive haloes, namely Milky Way, Andromeda and Triangulum galaxy.

We investigated the physical effects that lead to differences in the SFHs of such satellites, and found that the variety of star formation histories of satellite galaxies can be explained by the complex interplay between the satellite's evolution before and after its infall into the main host galaxy. In order to study the role played by the environment, the internal properties, and the orbit of each satellite in shaping the final SFH of these objects, we used a control sample of simulated isolated galaxies that cover the same stellar mass range as our selected satellites, to highlight any difference between the evolution of these two groups.

Satellites were selected to be more massive than $M_{vir} = 10^9 \mathrm{M}_\odot$ at infall time, and to contain more than 100 stellar particles at z=0, to ensure adequate resolution: this selection criteria provides a sample of 23 objects (Fig. 1). To determine whether the SF of a satellite has been suppressed or enhanced after infall, we assign each satellite galaxy to the corresponding isolated galaxy group according to their mass accretion history (Fig. 4), and compare the SFH of satellites after infall with the SFH that they would have had if they had kept evolving in isolation. We found, in agreement with previous studies, that the least luminous satellites stop forming stars well before infall, such that in this case the infall does not play any role in shaping their SFH (e.g. Benítez-Llambay et al. 2015, and references therein). More massive galaxies can, on the other hand, have their SF suppressed or enhanced if specific conditions are met. The main results of our work can be summarized as follows:

- after infalling into their respective hosts, notable differences can be found amongst the SFHs of satellites galaxies; while in the majority of the cases a suppression in star formation is observed, in about 25% of our sample a clear enhancement in star formation occurs (Fig. 1);
- the enhancement of SF after infall does not seem to correlate with the satellites' distance from host, infall time, total halo or stellar mass; it does, however, correlate with the amount of cold gas that a satellite galaxy had when infalling into its main host galaxy (Fig. 6), such that none of the satellites that infall with $M_{gas} \lesssim 10^7 M_\odot$ within their inner region (20% of $R_{vir}$) will form stars after infall;

- in addition to infalling with relatively large cold gas fractions ($M_{\rm gas,inf}(< 20\% R_{\rm vir})/M_{\rm vir,inf} \gtrsim 10^{-2}$ in our simulations, Fig. 7, left panel), another condition must be fulfilled in order for a satellite to be able to keep forming stars after infall, namely that its pericentric distance to the main host galaxy should be relatively large (average $d_{peri} \sim$47 kpc for our enhanced satellites, Fig. 7, right panel). This condition ensures that at pericentric passage the large fraction of existing gas is compressed rather than lost via ram-pressure stripping, so that the satellite is able to form new stars (Fig. 8 and Fig. 10);
- the peaks in star formation occuring after a satellite's infall correlate strongly with the satellite's orbit around its host, such that star formation is enhanced exactly at the time of pericentric passage, likely due to the compression of gas (Fig. 8); oppositely, satellites that fall in with large gas reservoir but have an extremely close-by approach with their hosts ($d_{peri} \sim$2-5 kpc, Fig. 9) will have their gas removed and their SF will be abruptly quenched;
- we found a case in which the SF of a satellite galaxy is affected not only by the satellite-host interaction, but also by the close-by passage with another dwarf galaxy, suggesting that this mechanism could also happen in the real Universe (Fig. 8, right-most panel);
- additionally, we identified clear peaks of SF in the host galaxies MW, M31 and M33, at the time their satellites reach pericenter, with this effect being more pronounced when multiple satellites approach the host at a similar time, a sign that satellite-host interactions not only affect the subsequent SFH of satellites, but they have a tangible impact on the final SFH of the host galaxy itself (Fig. 11);
- finally, when comparing the eccentricity and radial velocities of our simulated satellites with the observed Local Group ones, using the latest Gaia DR2 data, we observe that the enhanced satellites tend to be on more circular orbits with respect to quenched ones, and that some LG dwarfs follow similar trends (Fig. 12): in particular, we closely resemble the cases of Carina and Fornax, whose low eccentricity and large pericenters of 60 and 58 kpc, respectively, correlate with clear peaks in their SF.

Our simulations show that close encounters with the main host galaxy, as well as with other dwarfs, are a viable mechanism to explain the existence of multiple stellar populations in dwarf galaxies such as Carina, Leo I or Fornax (Hurley-Keller et al. 1998; Gallart et al. 1999; Monelli et al. 2003; Bono et al. 2010; Monelli et al. 2014; del Pino et al. 2013) In particular, we show an example of satellite galaxy whose SFH resembles the one of Carina: in Fig. 8, right most panel, we show how interactions with a dwarf first, and with the main host later, provoke two peaks of SF at intermediate and young ages, just as observed for the two SF peaks in Carina, which clearly indicate that such stars formed from gas with different abundance patterns (de Boer et al. 2014).

Moreover, recent work from Rusakov et al. (2021) also supports this scenario, since a clear correlation is observed between the intermediate and young SF events of Fornax dSphs and its pericentric passages around the Milky Way. Our study is also in agreement with very recent work by Ruiz-Lara et al. (2020), that show how the SFH of our own Milky Way galaxy could be impacted by the pericentric passage of its satellites: three narrow episodes of enhanced star formation have been derived for the MW, whose timing coincides very well with the proposed Sagittarius dwarf pericentric passage. Finally, in agreement with Miyoshi & Chiba (2020), who found a remarkable correlation between a smaller pericentric radius and a reduction of star formation after its peak, we find that a small pericenter leads to a removal of gas by ram-pressure and therefore to a suppression in SF: this effect has been also observed in Fillingham et al. (2019), who showed that quenching timescales are shorter for those dSphs with high orbital eccentricities.

A caveat to bear in mind is that some of our simulated satellites have, at z=0, more gas than the observed Local Group' dwarf spheroidals. Therefore, in the future, verifying our results with other simulations, run within different environment and with different initial conditions and sub-grid physics implementation, will strengthen the robustness of our conclusions.

Nevertheless, we have shown here that the quenching of SF in satellite galaxies after their infall into a main host galaxy is not the only possible outcome: enhanced star formation can also occur in satellite galaxies. This finding calls for a revision of galaxy formation processes currently implemented into semi-analytic models, which have historically taken into account only a crude quenching of satellite galaxies after their infall.

Ultimately, the complex interplay between the satellites' masses at reionization, their mass accretion history, gas reservoir at infall, type of orbit, evolution within a particular environment and close interactions with other galaxies, can give rise to a variety of SFHs reminiscent of the ones observed in the Local Group.

## ACKNOWLEDGEMENTS

ADC kindly thanks JFN and the University of Victoria for the hospitality during the preparation of this work. ADC is supported by a Junior Leader fellowship from 'La Caixa' Foundation (ID 100010434), fellowship code LCF/BQ/PR20/11770010. AK and RM are supported by the Ministerio de Ciencia, Innovación y Universidades (MICIU/FEDER) under research grant PGC2018-094975-C21. AK further thanks Matt Berninger for serpentine prison. The authors kindly acknowledge the PIs of the CLUES collaboration for data sharing, as well as A.Benítez-Llambay, C. Gallart, T. Ruiz-Lara and G. Battaglia for many useful discussions. We sincerely thank the referee for a careful review of our work.

## DATA AVAILABILITY

The data underlying this article will be shared on reasonable request to the corresponding author.